
\input harvmac
\magnification 1200 
\hsize=16.5truecm 
\vsize=23.5truecm 
\overfullrule=0pt
\parskip=3pt
 at 14truept
\overfullrule 0pt
\voffset -10pt 


\def\hexnumber@#1{\ifcase#1 0\or1\or2\or3\or4\or5\or6\or7\or8\or9\or
        A\or B\or C\or D\or E\or F\fi }

%
\def\boxit#1{\leavevmode\kern5pt\hbox{
	\vrule width.2pt\vtop{\vbox{\hrule height.2pt\kern5pt
        \hbox{\kern5pt{#1}\kern5pt}}
      \kern5pt\hrule height.2pt}\vrule width.2pt}\kern5pt}
\def\boxEq#1{\boxit{$\displaystyle #1$}}



\def\\{\hfil\break}
\def\la{\lambda}

\def\om{\omega}
\def\si{\sigma}
\def\Z{{\bf Z}}
\def\R{{\bf R}}
\def\l{\ell}

\def\b#1{\big(#1\big)}

\def\+{\oplus}
\def\x{\otimes}

\def\b#1{\kern-0.25pt\vbox{\hrule height 0.2pt\hbox{\vrule
width 0.2pt \kern2pt\vbox{\kern2pt \hbox{#1}\kern2pt}\kern2pt\vrule
width 0.2pt}\hrule height 0.2pt}}
\def\ST#1{\matrix{\vbox{#1}}}
\def\STrow#1{\hbox{#1}\kern-1.35pt}
\def\bv{\b{\phantom{1}}}

\font\huge=cmr10 scaled \magstep2
\font\small=cmr8


{\nopagenumbers
\rightline{hep-th/9612159}
\vskip 2cm
\centerline{{\huge\bf Demazure Characters and Affine Fusion Rules}}
\bigskip\bigskip\centerline{M.A.
Walton {\footnote{$^\dagger$}{\small Supported in part by NSERC. E-mail:
walton@hg.uleth.ca}}}
\bigskip

\centerline{{\it Physics Department,
University of Lethbridge}} \centerline{{\it Lethbridge, Alberta,
Canada\ \ T1K 3M4}}

\vskip 3cm \leftskip=2cm \rightskip=2cm
\noindent{{\bf Abstract:}}
The Demazure character formula is applied to the Verlinde formula for affine
fusion rules. We follow Littelmann's
derivation of a generalized  Littlewood-Richardson rule from Demazure
characters. A combinatorial rule for affine fusions does not result, however.
Only a modified version of the Littlewood-Richardson rule is
obtained that computes an (old) upper bound on the fusion coefficients of
affine $A_r$ algebras. We
argue that this is because the characters of simple Lie algebras appear in
this treatment, instead of the corresponding affine characters. The Bruhat
order on the affine Weyl group must be implicated in any combinatorial rule for
affine fusions; the Bruhat order on subgroups of this group (such as the finite
Weyl group) does not suffice.   \leftskip=0cm \rightskip=0cm

\vfill

\eject}

\pageno=1
\newsec{\bf Introduction}

The fusion coefficients $N_{l,m}^n\in \Z_{\ge0}$ of a rational conformal
field  theory (RCFT) may
be defined by\ \ref\Ver{Verlinde, E.: Fusion rules and modular transformations 
in 2D conformal field theory. Nucl.\ Phys.\ {\bf B300 [FS22]}, 360-376 (1988)} 
\eqn\Vrcft{{S_{\l,p}\over S_{0,p}}\ {S_{m,p}\over S_{0,p}}\ =\ 
\sum_{n\in\Phi}\ N_{l,m}^n\ {S_{n,p}\over S_{0,p}},\ \ \forall p\in \Phi.}
Here the indices $\l,m,n,p,\ldots\in\Phi$ label the primary fields of the RCFT,
with the index 0 specifying the identity field. $S_{i,j}$ denotes an element of
the matrix describing the transformation of
the characters of the primary fields under the change of torus modulus  
$\tau\rightarrow -1/\tau$.

We will restrict attention to those RCFTs realizing an affine Kac-Moody
algebra $X_{r,k}$ that is the central extension at fixed level $k$ of the loop
algebra of the simple Lie algebra $X_r$ of rank $r.$ These are often called
Wess-Zumino-Witten (WZW) models \ref\gw{Gepner, D., Witten, E.: Strings on
group  manifolds. Nucl. Phys. {\bf B278}, 493-549 (1986)}.

If the chiral algebra is not
extended, the primary fields of such theories are in one-to-one
correspondence with the integrable highest-weight representations of
$X_{r,k}.$ The set of highest weights of these affine representations can
therefore be used to label the primary fields. Alternatively, we can label the
primary fields using the following set of dominant weights of $X_r$:
\eqn\Pplusk{P_\ge (X_{r,k})\ :=\ \left\{\ \la=\sum_{i=1}^r \la_i\om^i\ \vert\
\la_i\in\Z_{\ge0},\ \sum_{i=1}^r \la_ia_i^\vee\le k\ \right\}\ \ .} 
$\om^i$ denotes the $i$-th fundamental weight of $X_r$, and $a_i^\vee$ is the
corresponding co-mark, so that $1+\sum_{i=1}^ra_i^\vee=h^\vee,$ the dual
Coxeter number. $F(X_r)$ will stand for the set of fundamental weights of
$X_r$. The fusion rules of WZW models may be written as
\eqn\Vwzw{{S_{\la,\zeta}\over S_{0,\zeta}}\ {S_{\mu,\zeta}\over S_{0,\zeta}}\
=\  \sum_{\nu\in\ P_\ge (X_{r,k})}\ N_{\la,\mu}^\nu\ {S_{\nu,\zeta}\over
S_{0,\zeta}},\ \ \ \ \ \forall \zeta\in P_\ge (X_{r,k}).} For short, we will
refer to these as affine fusion rules.

The ratios of matrix elements of $S$ appearing in \Vwzw\ are
intimately related to the characters of representations of $X_r$ \ref\KP{Kac,
V.\ G., Peterson, D.: Infinite-dimensional Lie algebras, theta functions and
modular forms. Adv.\ Math.\ {\bf 53}, 125-264 (1984)}.  Let $R(\la)$
be the integrable representation of $X_r$ with highest weight $\la$, and let
$P(\la)$ denote the set of its weights. If 
\eqn\Pint{P(X_r)\ :=\ \left\{\ \la=\sum_{i=1}^r \la_i\om^i\ \vert\
\la_i\in\Z\ \right\}\ \ }
is the set of integral weights of $X_r$, we have $P(\la)\subset P(X_r)$. Let
$e^\lambda$ denote the formal exponential of the weight $\lambda$, with the
property $e^\lambda e^\mu = e^{\lambda+\mu}$. The formal character of $R(\la)$
may be written as  \eqn\chmult{{\rm ch}_\la\ :=\ \sum_{\si\in P(\la)} {\rm
mult}_\la(\si)\  e^\sigma\ \ ,} where ${\rm mult}_\la(\si)$ is the multiplicity
of the weight $\si$ in the representation $R(\la)$, and
$\rho=\sum_{i=1}^r\om^i$ is the Weyl vector. 

Define \eqn\eval{e^\si(\zeta)\ :=\ \exp \left[{{-2\pi
i}\over{k+h^\vee}}\si\cdot(\zeta+\rho)\right]\ \ .} 
Kac and  Peterson showed that
\eqn\Sch{ {S_{\la,\zeta}\over S_{0,\zeta}}\ =\ {\rm ch}_\la(\zeta)\ ,\ \ \
(\ \la,\zeta\in P_\ge (X_{r,k})\ )\ .}

This remarkable result allows us to relate the affine fusion coefficients
$N_{\la,\mu}^\nu$ to the coefficients appearing in the decomposition of a
tensor product of two representations of $X_r$. Let
\eqn\Pplus{P_\ge (X_r)\ :=\ \left\{\ \la=\sum_{i=1}^r \la_i\om^i\ \vert\
\la_i\in\Z_{\ge0}\ \right\}\ \ }
be the set of dominant weights of $X_r$. Then, if 
the tensor product
coefficients $T_{\la,\mu}^\nu$ are defined by
\eqn\tp{R(\la)\otimes R(\mu)\ =\ \bigoplus_{\nu\in P_\ge (X_r)}\ T_{\la,\mu}^\nu\
R(\nu)\ \ ,} the corresponding characters obey
\eqn\chtp{{\rm ch}_\la\ {\rm ch}_\mu\ =\ \sum_{\nu\in P_\ge (X_r)}
\ T_{\la,\mu}^\nu\ {\rm ch}_\nu\ \ .} 
The Weyl character formula is
\eqn\Weyl{ {\rm ch}_\la\ =\ { {\sum_{w\in W} ({\rm det}w)\
e^{w(\la+\rho)}} \over {\sum_{w\in W} ({\rm det}w)\
e^{w\rho}} }\ \ ,}
where $W$ is the Weyl group of $X_r$. Therefore, 
\eqn\Weylz{ {\rm ch}_\la(\zeta)\ =\ { {\sum_{w\in W} ({\rm det}w)\
\zeta^{w(\la+\rho)}} \over {\sum_{w\in W} ({\rm det}w)\
\zeta^{w\rho}} }\ \ ,} where we
have defined
\eqn\notation{\zeta^\si\ :=\ e^\si(\zeta)\ \ }
for notational convenience. The Weyl formula makes manifest the Weyl symmetry of
the characters:
\eqn\Wsymm{ {\rm ch}_\la(\zeta)\ =\ ({\rm det}w)\ {\rm ch}_{w.\la}(\zeta)\ =\ 
{\rm ch}_{\la}(w.\zeta)\ \ \ ,\ \ \ \ \forall w\in W\ (\zeta\in P_{\bf
R}(X_r))\ ,} where $w.\la:=w(\la+\rho)-\rho$ denotes the shifted action of the
Weyl group element $w$, and $P_\R(X_r):=\{\ \la=\sum_{i=1}^r
\la_i\om^i\ \vert\ \la_i\in\R\ \}$. When $\zeta\in P(X_{r})$, this
symmetry can be extended to include the elements of the Weyl group of
$X_{r,k}$, the so-called affine Weyl group $\widehat W$:
\eqn\aWsymm{ {\rm ch}_\la(\zeta)\ =\ ({\rm det}w)\ {\rm ch}_{w.\la}(\zeta)\ =\ 
{\rm ch}_{\la}(w.\zeta)\ \ \ ,\ \ \forall w\in \widehat W\ \ 
(\ \zeta\in P(X_{r})\ ).}
Let $S(X_r)$ signify the set of simple roots of $X_r$. $W$ is generated by the
primitive reflections $r_\alpha$,   \eqn\ralpha{r_\alpha\ \la\ :=\ \la\ -\
(\la\cdot\alpha^\vee)\ \alpha\ ,\ \ \alpha\in S(X_r),}
where $\alpha^\vee$ is the simple co-root dual to the simple root $\alpha$. 
$\widehat W$ has one additional generator $r_{\alpha_0}$ corresponding to the
0-th affine simple root $\alpha_0=\delta-\theta$ \ref\Kac{Kac, V.\ G.: Infinite 
dimensional Lie algebras, 3rd edition. 
Cambridge: Cambridge University Press 1990}. Here $\theta$
is the highest root of $X_r$, and $\delta$ is the imaginary root of the
affine algebra that yields $\delta\cdot\lambda=k$ for any level $k$
weight $\lambda$ (and $\delta\cdot\beta=0$ for any root $\beta$). The action of
$r_{\alpha_0}$ in \aWsymm\ is  therefore \eqn\rzero{r_{\alpha_0}.\la\ =\
r_\theta.\la +(k+h^\vee)(\theta-\delta)\ \ .}
Since $\zeta$ corresponds to a level $k$ weight, and $\rho$ to a level
$h^\vee$ one, we have $\zeta^\delta=1$. Then the identity \eqn\zrzero
{\zeta^{r_{\alpha_0}.\la}\ =\ \zeta^{r_\theta.\la
+(k+h^\vee)\theta}\ =\ \zeta^{r_\theta.\la}}
leads to the full affine Weyl invariance of \aWsymm.

Comparing equations \Vwzw\ and \chtp, taking into account the Kac-Peterson
relation \KP\ and the affine Weyl symmetry \aWsymm, one finds \ref\MW{Walton, M.A.: Algorithm for 
WZW fusion rules: a proof. Phys.
Lett. {\bf 241B}, 365-368 (1990); Fusion rules in Wess-Zumino-Witten models.
Nucl. Phys. {\bf B340}, 777-790 (1990)}\Kac\ref\FGP{Furlan, P., Ganchev, A.,
Petkova, V.: Quantum groups and fusion rules multiplicities. Nucl.\ Phys.\ {\bf
B343}, 205-227 (1990)}\ref\GoodW{Goodman, F.M., Wenzl, H.:
Littlewood-Richardson coefficients for Hecke algebras at roots of unity. Adv.\
Math.\ {\bf 82}, 244-265 (1990)}: \eqn\NT{N_{\la,\mu}^\nu\ =\ \sum_{w\in\widehat W}\
({\rm det}w)\ T_{\la,\mu}^{w.\nu}\ \ .} Using the well-known connection between
multiplicities and tensor product coefficients, \eqn\Tmult{T_{\la,\mu}^\nu\ =\
\sum_{w\in W}\ ({\rm det}w)\ {\rm mult}_\mu(w.\nu-\lambda)\ ,}
we can also write
\eqn\Nmult{N_{\la,\mu}^\nu\ =\ \sum_{w\in\widehat W}\
({\rm det}w)\ {\rm mult}_\mu(w.\nu-\lambda)\ \ .}
These formulas make it clear that the affine
fusion coefficients are integers, while it is not at all evident from the form
of the matrix $S$ in \Vwzw. They do not, however, confirm that the fusion
coefficients are  non-negative integers, as must be. 

That the tensor product coefficients are non-negative is made manifest by 
rules for their computation that reduce to {\it counting} certain objects,
usually Young tableaux (see \ref\BKW{Black, G.R.E., King, R.C., Wybourne, B.G.:
Kronecker products for compact semisimple Lie groups. J. Phys. A: Math. Gen.
{\bf 16}, 1555-1589 (1983)}, for example). The most famous is the
Littlewood-Richardson rule that computes the tensor product coefficients of the
simple Lie algebra $A_r$. Such rules do not involve over-counting and
cancellations, and will be called combinatorial rules. 

Algorithms like the 
Littlewood-Richardson rule involve Young tableaux. 
The action of the affine Weyl group on the Young tableaux for classical
simple Lie algebras can be simply described. So, \NT\ allows the
combinatorial methods for computing tensor products to be adapted to the
computation of affine fusion rules \ref\CK{Cummins, C.J., King, R.C.: WZW
fusion rules for the classical Lie algebras. Can. J. Phys. {\bf 72}, 342-344
(1994)}. The adaptation is not combinatorial, however; the factor ${\rm
det}w=\pm 1$ in \NT\ means the adapted methods involve over-counting and
cancellations. 

Can one find a purely combinatorial method for the computation of affine fusion
rules? To the best of our knowledge, no such rule is known\foot{Of course,
for sufficiently  large level $k$, when
the fusion coefficient coincides with the corresponding tensor product
coefficient, we can use the combinatorial rules for tensor products.}. We
believe that such a rule, if it exists, would find wide application, much as the
original  Littlewood-Richardson rule has. More importantly, however, it would
provide a constructive mathematical proof that fusion coefficients are
non-negative integers, something that is obvious and necessary from a physical
point of view. It may also deepen our understanding of affine fusions; perhaps
it will provide a definition of affine fusion expressed solely in terms of
representations, analogous to that for simple Lie algebra tensor products. 

We may try to use \NT\ as a guide to a combinatorial rule. The problem with
\NT, however, is the factor ${\rm det}w=\pm 1$, inherited from the Weyl
character formula \Weyl. So, perhaps a non-negative character formula can lead
to a combinatorial rule. The Demazure character formula
\ref\Dem{Demazure, M.: D\'esingularisation des vari\'et\'es de Schubert
g\'en\'eralis\'ees. Ann. Ecole Norm. Sup. {\bf 7}, 53-88
(1974);\ Une nouvelle formule des caract\`eres. Bull. Sc.
Math., $2^{\rm e}$ s\'erie, {\bf 98}, 163-172 (1974)}\ref\Kum{Kumar, S.:
Demazure character formula in arbitrary Kac-Moody setting. Invent.\ Math.\ {\bf
89}, 395-423 (1987)} is such a non-negative formula, and Littelmann
\ref\Lit{Littelmann, P.: A generalization of the Littlewood-Richardson rule. J.
Algebra {\bf 130}, 328-368 (1990)} has derived the Littlewood-Richardson rule
and generalizations from it. In this work, we adapt his methods to the
computation of affine fusion rules, in an attempt to find a purely
combinatorial rule. Although our results are only valid for $X_r=A_r$, we 
use universal language appropriate to all simple Lie algebras (and their
untwisted affine analogues) whenever possible. This reflects our hope that
a universal result, valid for all $X_{r,k}$, will eventually be found. 

As far as we know, this work is the first direct application of Demazure
character formulas to affine fusions. In light of Littelmann's results, this
is a natural attempt at finding a combinatorial rule for affine
fusions. Unfortunately, no such rule is obtained. Our effort does make it clear
why, however. We hope it will therefore help as a stepping stone to a
combinatorial rule for affine fusions.

\newsec{\bf Demazure Character Formulas }

Introduce the primitive Demazure operators $D_\alpha$:
\eqn\Dalpha{D_\alpha(e^\mu)\ :=\ { {e^{\mu}-e^{r_\alpha.\mu}} \over 
{1 - e^{-\alpha}} }\ \ ,}
for all $\alpha\in S(X_r)$. These linear 
operators are associated with the primitive reflections of the Weyl group $W$
of $X_r$, and are defined for all $\zeta\in P_\ge(X_r)$, and $\mu\in P(X_r)$.
More explicitly,  \eqn\Dalphalong{D_\alpha(e^\mu)\ =\ \cases{ 
e^\mu+\ldots+e^{\mu-(\mu\cdot\alpha^\vee)\alpha}\ &,\
$\mu\cdot\alpha^\vee\ge0$\ ;\cr
0\ \ \ \ \ \ \ \ \ \ \ \ \ \ &,\ $\mu\cdot\alpha^\vee=-1$\ ;\cr
-e^{\mu+\alpha}-\ldots-e^{\mu+(-\mu\cdot\alpha^\vee-1)\alpha}\
&,\ $\mu\cdot\alpha^\vee\le-2$\ \ .\cr} }
A Demazure operator can also be associated to every element of the Weyl group
of $X_r$. Suppose $w=r_{\beta_1}r_{\beta_2}\cdots r_{\beta_\l}$, with all
$\beta_i\in S(X_r)$, is a  decomposition of $w\in W$. Such a decomposition is
said to have length $\ell$. For fixed $w\in W$, any decomposition of minimum
length is called a reduced decomposition. The length $\ell(w)$ of any reduced
decomposition of $w\in W$ is known as the length of $w$. Now, if 
$w=r_{\beta_1}r_{\beta_2}\cdots r_{\beta_\l}$ is a reduced decomposition of
$w\in W$,  then we define  \eqn\Dw{D_w\ :=\ D_{\beta_1}\circ
D_{\beta_2}\circ\cdots\circ D_{\beta_\l}\ \ .} It makes sense to label these
operators with the Weyl group element $w$, since they do not depend on which
reduced decomposition is used. It is important, however, that a {\it reduced}
decomposition be used in the definition of $D_w$; we have $r_\alpha^2=id$, but 
\eqn\Disquare{ D_\alpha\left(\ D_\alpha(e^\mu)\right)\ =\
D_\alpha(e^\mu)\ \ .} This last relation follows from the more general one:
\eqn\DaDa{D_\alpha \left(\ e^\la\ D_\alpha(e^\mu)\ \right)\ =\
D_\alpha(e^\la)\ D_\alpha(e^\mu)\ \ .}

The Weyl group of any simple Lie algebra contains a unique element of maximum
length. If $w_L$ symbolizes the longest element of the Weyl group $W$, we have
the following Demazure character formula: \eqn\Dch{{\rm ch}_\la\ =\
D_{w_L} (e^\la)\ \ .}

It is also useful to introduce other primitive Demazure operators related to the
$D_\alpha$:
\eqn\dalpha{d_\alpha(e^\mu)\ :=\ (D_\alpha-id)(e^\mu)\ =\ {
{e^{\mu-\alpha}-e^{r_\alpha.\mu}} \over  {1 - e^{-\alpha}} }\ \ ,}
for all $\alpha\in S(X_r)$. More explicitly, 
\eqn\dalphalong{d_\alpha(e^\mu)\ =\ \cases{ 
e^{\mu-\alpha}+\ldots+e^{\mu-(\mu\cdot\alpha^\vee)\alpha}\ &,\
$\mu\cdot\alpha^\vee\ge1$\ ;\cr
0\ \ \ \ \ \ \ \ \ \ \ \ \ \ &,\ $\mu\cdot\alpha^\vee=0$\ ;\cr
-e^{\mu}-\ldots-e^{\mu-(\mu\cdot\alpha^\vee+1)\alpha}\
&,\ $\mu\cdot\alpha^\vee\le-1$\ \ .\cr} }
These operators can also be generalized to $d_w$, for each $w\in
W\backslash id$, in a way similar to that defining the $D_w$. For later
convenience, we also define $d_{id}(e^\la):=e^\la$.

To write other Demazure character formulas, we need the Bruhat partial order
on the elements of the Weyl group $W$. Let $r_\alpha$ be the Weyl
reflection across the hyperplane in weight space normal to the root
$\alpha$, i.e. $r_\alpha \la=\la-(\la\cdot\alpha^\vee)\alpha$. The length
$\l(w)$ of a Weyl group element $w$ is defined as the number of primitive
reflections in a reduced decomposition of the element. For $v_1,\
v_2\in W$, we write $v_1\leftarrow v_2$ if and only if both $v_1=r_\alpha v_2$
for some positive root $\alpha$, and $\l(v_1)=\l(v_2)+1$. Then for $w,\tilde
w\in W$, the Bruhat order is defined by $w\succ\tilde w$ if $w\leftarrow
v_1\leftarrow v_2\leftarrow\cdots\leftarrow v_n\leftarrow\tilde w$, for some
$v_1,v_2,\ldots,v_n\in W$. 

Another formula for the character of the representation $R(\la)$ is
\eqn\dch{{\rm ch}_\la\ =\ \sum_{v\in W}\ d_v(e^\la)\ \ .}
A generalization is 
\eqn\dchw{D_w(e^\la)\ =\ \sum_{v\preceq w}\ d_v(e^\la)\ \ .}
The formula 
\eqn\Dd{D_\alpha\left(d_{r_\alpha v}(e^\mu)\right)\ =\
d_{r_\alpha v}(e^\mu)\ +\ d_v(e^\mu)\ ,\ \ {\rm if}\ r_\alpha
v\prec v\ ,\ \ \alpha\in S(X_r)\ \ ,}  makes explicit the relation between the
two types of Demazure operators. Using \dalpha\ this translates into
\eqn\dandd{d_v (e^\mu)\ =\ d_\alpha\circ d_{r_\alpha v} (e^\mu)\ ,\ \
{\rm if}\ r_\alpha v\prec v\ ,\ \ \alpha\in S(X_r)\ \ .}

\newsec{\bf Tableaux and the Weyl Group}

In order to relate the Demazure formulas to the Littlewood-Richardson rule, 
the Weyl group must be related to 
Young tableaux and standard tableaux. This is done by recognizing
certain sequences of elements of $W$ as the appropriate generalizations of
standard tableaux. The 
so-called minimal defining chains (MDCs) of Weyl group elements are in
one-to-one correspondence with certain vectors of fixed weight that form
a basis of the irreducible representation $R(\la)$ of $X_r$ \ref\LS{Lakshmibai, V., Seshadri, C.S.: Standard
monomial theory. In: Proceedings of the Hyderabad Conference on Algebraic
Groups, pp. 279-323. Madras: Manoj Prakashan 1991}\ref\Lpaths{Littelmann, P.: A Littlewood-Richardson rule for symmetrizable
Kac-Moody algebras. Inv. Math. {\bf 116}, 329-346 (1994)}. For the algebra
$A_r$, it is well known that such vectors are in one-to-one correspondence
with standard tableaux. MDCs are what work for all simple Lie algebras $X_r$.
 
Suppose $\mu$ is the highest weight
of an integrable representation $R(\mu)$ of $X_r$. Such a weight can be
expressed as a sum of fundamental weights $\mu=\nu^1+\nu^2+\ldots+\nu^n$, where
$\nu^j\in F(X_r),\ 1\le j\le n.$ If we fix an order $>$ on the elements of
$F(X_r)$, so that we can also impose $\nu^i\ge \nu^{i+1}$ on the sum, the
expression is unique. Now consider a sequence $c=(w_1,w_2,\ldots,w_n)$ of $n$
Weyl group elements $w_i\in W$. The
sequence, or chain, is a defining chain if it respects the Bruhat order:
$w_1\preceq w_2\preceq\cdots\preceq w_n$. A sequence of weights of fundamental
representations corresponding to each defining chain and weight $\mu$ is given
by \eqn\seq{P_\mu(c)\ :=\ (w_1\nu^1,w_2\nu^2,\ldots,w_n\nu^n)\ \ .}
Different defining chains, however, lead to the same weight-sequence. A unique
{\it minimal defining chain} (MDC) can be associated with
the weight sequence: $c=(w_1,\ldots,w_n)$ is a minimal defining chain if
for any other defining chain $\tilde c=(\tilde
w_1,\ldots,\tilde w_n)$ satisfying $P_\mu(c)=P_\mu(\tilde c)$, we have 
$w_j\preceq\tilde w_j$, for all $1\le j\le n$.

Define the $\mu$-weight $p_\mu[c]$ of the MDC $c$ as
\eqn\plac{p_\mu[c]\ =\ p_\mu\left[(w_1,\ldots,w_n)\right]\ :=\
w_1\nu^1+\ldots+w_n\nu^n\ .} Let ${\cal C}_\mu[\sigma]$ be the set of MDCs
with $\mu$-weight $\sigma$, and let ${\cal C}_\mu(w)$ denote the set of MDCs
with last element equal to $w\in W$. The weights of a representation $R(\mu)$
and those of MDCs are related: \eqn\multc{{\rm mult}_\mu(\sigma)\ =\ \vert{\cal
C}_\mu[\sigma]\vert\ \ ,  }
so that 
\eqn\chc{{\rm ch}_\mu\ =\ \sum_{\sigma\in P(\mu)}\ \sum_{c\in {\cal
C}_\mu[\sigma]}\ e^{p_\mu[c]}\ \ .} 
Also relevant here is the connection with
Demazure operators: \eqn\dC{d_w(e^\lambda)\ =\ \sum_{c\in {\cal
C}_\lambda(w)}\ e^{p_\lambda[c]}\ \ .}

In the case $X_r=A_r$, the usual standard tableaux are recovered as
follows. Let $\{\ e_i\ |\ i=1,\ldots,r+1;\ e_i\cdot e_j=\delta_{i,j}\ \}$
be an orthonormal basis of ${\bf R}^{r+1}$. The fundamental weights can be
chosen to be $\omega^i=e_1+e_2+\ldots+e_i-i\psi/(r+1),$ with
$\psi=\sum_{i=1}^{r+1} e_i$ \ref\bou{Bourbaki, N.: Groupes et alg\`ebres de
Lie, Chapitres IV-VI. Paris: Hermann 1968}. The Weyl group of $A_r$ acts as the
group of permutations on the $e_i$. Therefore the weights $w_j\nu^j$ $(w_j\in
W,\ \nu^j\in F(A_r))$ in the weight sequence $P_\mu(c)=
(w_1\nu^1,w_2\nu^2,\ldots,w_n\nu^n)$ of a MDC $c$ can be associated with a
column of boxes, with numbers from 1 to $r+1$ in each, increasing down the
column. A box $\ST{\STrow{\b2}}$, say, corresponds to a summand $e_2$ in the
expression for the weight, modulo multiples of $\psi$. If the fundamental
weights of $A_r$ are ordered $\omega^1<\omega^2<\cdots<\omega^r$, then if the
columns of boxes corresponding to the weights in the weight-sequence of a MDC
are assembled into a tableaux, the columns will have heights that do not
increase from left to right. Furthermore, the Bruhat order imposed on a MDC
ensures that the numbers in the rows of the tableau also do not increase from
left to right. But these  are precisely the defining properties of a standard
tableau\foot{These are sometimes called semi-standard tableaux, when
standard tableaux designate those with numbers increasing in both rows and
columns.} for $A_r$. (The further {\it minimal} property of the MDCs is
necessary for  \multc\ and \dC.) 

The MDCs, corresponding weight sequences and standard
tableaux associated with vectors in the $A_2$ representation $R(2,1)$ of highest
weight $2\om^1+\om^2=:(2,1)$ are given as an example, in Table 1. There we
use the shorthand notation $r_{\alpha_i}=:r_i$.

\topinsert
\centerline{
{\vbox{\offinterlineskip
\hrule
\halign{& \vrule# & \strut \quad \hfill#\hfill \quad
& \vrule# & \strut \quad \hfill#\hfill \quad
& \vrule# & \strut \quad \hfill#\hfill \quad
& \vrule# & \strut \quad \hfill#\hfill \quad & \vrule# \cr
height3pt & \omit && \omit && \omit && \omit &\cr
& MDC && weight\ sequence && tableau && weight  & \cr
height3pt & \omit && \omit && \omit && \omit &\cr
\noalign{\hrule}
height2pt & \omit && \omit && \omit && \omit &\cr
\noalign{\hrule}
height4pt & \omit && \omit && \omit && \omit &\cr
&$(id,id,id)$&&$((0,1),(1,0),(1,0))$&&
$\ST{\STrow{\b1\b1\b1}\STrow{\b2}}$
&&$(2,1)$ &\cr
height3pt & \omit && \omit && \omit && \omit &\cr
\noalign{\hrule}
height4pt & \omit && \omit && \omit && \omit &\cr
&$(id,id,r_1)$&&$((0,1),(1,0),(-1,1))$&&
$\ST{\STrow{\b1\b1\b2}\STrow{\b2}}$
&&$(0,2)$
&\cr height3pt & \omit && \omit && \omit && \omit &\cr
\noalign{\hrule}
height4pt & \omit && \omit && \omit && \omit &\cr
&$(id,r_1,r_1)$&&$((0,1),(-1,1),(-1,1))$&&
$\ST{\STrow{\b1\b2\b2}\STrow{\b2}}$
&&$(-2,3)$
&\cr height3pt & \omit && \omit && \omit && \omit &\cr
\noalign{\hrule}
height4pt & \omit && \omit && \omit && \omit &\cr
&$(r_2,r_2,r_2)$&&$((1,-1),(1,0),(1,0))$&&
$\ST{\STrow{\b1\b1\b1}\STrow{\b3}}$
&&$(3,-1)$
&\cr height3pt & \omit && \omit && \omit && \omit &\cr
\noalign{\hrule}
height4pt & \omit && \omit && \omit && \omit &\cr
&$(id,id,r_2r_1)$&&$((0,1),(1,0),(0,-1))$&&
$\ST{\STrow{\b1\b1\b3}\STrow{\b2}}$
&&$(1,0)$
&\cr height3pt & \omit && \omit && \omit && \omit &\cr
height4pt & \omit && \omit && \omit && \omit &\cr
&$(r_2,r_2,r_1r_2)$&&$((1,-1),(1,0),(-1,1))$&&
$\ST{\STrow{\b1\b1\b2}\STrow{\b3}}$
&&$(1,0)$
&\cr height3pt & \omit && \omit && \omit && \omit &\cr
\noalign{\hrule}
height4pt & \omit && \omit && \omit && \omit &\cr
&$(r_2,r_1r_2,r_1r_2)$&&$((1,-1),(-1,1),(-1,1))$&&
$\ST{\STrow{\b1\b2\b2}\STrow{\b3}}$
&&$(-1,1)$
&\cr height3pt & \omit && \omit && \omit && \omit &\cr
height4pt & \omit && \omit && \omit && \omit &\cr
&$(id,r_1,r_2r_1)$&&$((0,1),(-1,1),(0,-1))$&&
$\ST{\STrow{\b1\b2\b3}\STrow{\b2}}$
&&$(-1,1)$
&\cr height3pt & \omit && \omit && \omit && \omit &\cr
\noalign{\hrule}
height4pt & \omit && \omit && \omit && \omit &\cr
&$(r_1r_2,r_1r_2,r_1r_2)$&&$((-1,0),(-1,1),(-1,1))$&&
$\ST{\STrow{\b2\b2\b2}\STrow{\b3}}$
&&$(-3,2)$
&\cr height3pt & \omit && \omit && \omit && \omit &\cr
\noalign{\hrule}
height4pt & \omit && \omit && \omit && \omit &\cr
&$(r_2,r_2,r_2r_1)$&&$((1,-1),(1,0),(0,-1))$&&
$\ST{\STrow{\b1\b1\b3}\STrow{\b3}}$
&&$(2,-2)$
&\cr height3pt & \omit && \omit && \omit && \omit &\cr
\noalign{\hrule}
height4pt & \omit && \omit && \omit && \omit &\cr
&$(r_2,r_1r_2,r_2r_1r_2)$&&$((1,-1),(-1,1),(0,-1))$&&
$\ST{\STrow{\b1\b2\b3}\STrow{\b3}}$
&&$(0,-1)$
&\cr height3pt & \omit && \omit && \omit && \omit &\cr
height4pt & \omit && \omit && \omit && \omit &\cr
&$(id,r_2r_1,r_2r_1)$&&$((0,1),(0,-1),(0,-1))$&&
$\ST{\STrow{\b1\b3\b3}\STrow{\b2}}$
&&$(0,-1)$
&\cr height3pt & \omit && \omit && \omit && \omit &\cr
\noalign{\hrule}
height4pt & \omit && \omit && \omit && \omit &\cr
&$(r_1r_2,r_1r_2,r_2r_1r_2)$&&$((-1,0),(-1,1),(0,-1))$&&
$\ST{\STrow{\b2\b2\b3}\STrow{\b3}}$
&&$(-2,0)$
&\cr height3pt & \omit && \omit && \omit && \omit &\cr
\noalign{\hrule}
height4pt & \omit && \omit && \omit && \omit &\cr
&$(r_2,r_2r_1,r_2r_1)$&&$((1,-1),(0,-1),(0,-1))$&&
$\ST{\STrow{\b1\b3\b3}\STrow{\b3}}$
&&$(1,-3)$
&\cr height3pt & \omit && \omit && \omit && \omit &\cr
\noalign{\hrule}
height4pt & \omit && \omit && \omit && \omit &\cr
&$(r_1r_2,r_2r_1r_2,r_2r_1r_2)$&&$((-1,0),(0,-1),(0,-1))$&&
$\ST{\STrow{\b2\b3\b3}\STrow{\b3}}$
&&$(-1,-2)$
&\cr height3pt & \omit && \omit && \omit && \omit &\cr
\noalign{\hrule}
}}}}

\smallskip
\leftskip=1cm
\rightskip=1cm
\noindent
\baselineskip=12pt
{\bf Table 1.} Minimal defining chains and corresponding standard 
tableaux for the $A_2$ representation of highest weight
$2\omega^1+\omega^2=:(2,1).$ \bigskip
\leftskip=0cm
\rightskip=0cm
\baselineskip=13pt
\endinsert

If a standard tableau
corresponds to a weight-sequence $P_\mu(c)$ of the MDC $c$, then we say that
it has shape $\mu$. The shape of a standard tableau is determined by the
configuration of its boxes, and not by the numbers therein. If the numbers are
removed from a standard tableau of shape $\mu$, a Young tableau of
shape $\mu$ results. Young tableaux also enter into the Littlewood-Richardson
rule.

It will be useful to define subchains obtained from other chains by keeping
the last $i$ elements on the right, and also by keeping the first $n-i$
elements on the left. If $c=(w_1,w_2,\ldots,w_n)$ is a defining chain, we
define \eqn\RLc{R_i(c)\ :=\ (w_{n-i+1},w_{n-i+2},\ldots,w_n)\
,\ {\rm and}\ \ L_i(c)\ :=\ (w_1,w_2,\ldots,w_{n-i})\ \ .} If
$\mu=\nu^1+\nu^2+\ldots+\nu^n\in P_\ge (X_r),$ we also define
\eqn\RLmu{R_i(\mu)\ :=\ \nu^{n-i+1}+\nu^{n-i+2}+\ldots+\nu^n\ ,\ {\rm and}\ \
L_i(\mu)\ :=\ \nu^1+\nu^2+\ldots+\nu^{n-i}\ \ .}

\newsec{\bf Littelmann's Generalization of the Classical Littlewood-Richardson
Rule}

From \chtp\ and the Weyl character formula \Weyl, we have
\eqn\preth{\sum_{w\in W} (\det w) e^{w(\lambda+\rho)} {\rm ch}_\mu\
=\ \sum_{\nu\in P_\ge (X_r)} T_{\lambda,\mu}^\nu\ \sum_{w\in W} (\det w)
e^{w(\nu+\rho)}\ .}
Defining the linear operator $\Theta$ by
\eqn\Th{\Theta(e^\lambda)\ :=\ \sum_{w\in W}(\det w)
e^{w(\lambda+\rho)}\ \ ,}
this can be rewritten as
\eqn\Thi{\Theta\left(e^\lambda{\rm ch}_\mu\right)\ =\
\Theta\left(\sum_{\nu\in P_\ge (X_r)} T_{\lambda,\mu}^\nu e^\nu\right)\ .}

What Littelmann showed was that the left hand side simplifies because it
contains terms of the form 
\eqn\thdzero{\Theta\left( e^\kappa d_v(e^\gamma) \right)\ =\ 0,\ \ \ {\rm
for}\  r_\alpha v\prec v\in W,\ \kappa\cdot\alpha^\vee=0,\ \ \alpha\in S(X_r)\
.} To see why these contributions vanish, first note that by the definition \Th,
we get 
\eqn\Thw{\Theta\left(e^{w.\gamma}\right)\ =\ (\det w)\
\Theta\left(e^\gamma\right)\ ,\ \ \ \forall w\in W\ \ ,}
so that 
\eqn\ThD{\Theta\left( D_\alpha(e^\gamma)\right)\ =\
\Theta\left(e^\gamma\right)\ ,\ \ \ \forall \alpha\in S(X_r)\ \ .}
Eqn. \DaDa\ above then gives
\eqn\ThDaDa{\Theta\left(e^\kappa D_\alpha(e^\gamma)\right)\ =\ 
\Theta\left(D_\alpha(e^\kappa D_\alpha(e^\gamma))\right)\ =\ 
\Theta\left(D_\alpha(e^\kappa) D_\alpha(e^\gamma)\right)\ =\ 
\Theta\left(D_\alpha(e^\kappa) e^\gamma\right)\ ,}
for any simple root $\alpha\in S(X_r)$. This implies
\eqn\Thdada{\Theta\left(e^\kappa d_\alpha(e^\gamma)\right)\ =\ 
\Theta\left(d_\alpha(e^\kappa) e^\gamma\right)\ ,}
by the definition \dalpha\ of the operators $d_\alpha$. 
From \dandd\ we then get
\eqn\Thdz{\Theta\left( e^\kappa d_v(e^\gamma) \right)\ =\ 
\Theta\left( d_\alpha(e^\kappa) d_{r_\alpha v}(e^\gamma) \right)\ \
,\ {\rm for}\  r_\alpha v\prec v\ .} 
But since $d_\alpha(e^\kappa)=0$ for $\kappa\cdot\alpha^\vee=0$, \thdzero\
is proved. 

The vanishing contributions to the tensor product $R(\lambda)\otimes R(\mu)$
encoded in \thdzero\ can be identified easily using the MDCs. From \chc, the
left hand side of \Thi\ contains expressions of the form 
\eqn\colbycol{\Theta\left(e^{\lambda+p_\mu[c]}\right)\ =\ 
\Theta\left(e^{[ \lambda+p_{R_i(\mu)}\left[R_i(c)\right]\
]+p_{L_i(\mu)}\left[L_i(c)\right]}\right)\ \ ,   } where $1\le i\le n$. 
Suppose that for the particular MDC under consideration 
\eqn\iplus{\eqalign{\lambda+p_{R_i(\mu)}\left[R_i(c)\right]&\in P_\ge (X_r),\ {\rm
for}\ i=1,\ldots,\ell<n\ ;\cr
\lambda+p_{R_{\ell+1}[\mu]}&(R_{\ell+1}(c))\not\in P_\ge (X_r)\ .\cr}}
For $X_r=A_r$, we have $w\nu\cdot\alpha^\vee\in\{-1,0,1\}$, for all
$\alpha\in S(X_r)$ and $\nu\in F(X_r)$. With $c=(w_1,\ldots,w_n)$ and
$P_\mu(c)=(w_1\nu^1,\ldots,w_n\nu^n)$, \iplus\ implies that  
$\left(\lambda+p_{R_\ell(\mu)}\left[R_\ell(c)\right]\right)\cdot\alpha^\vee=0$
and 
$\left(\lambda+p_{R_{\ell+1}(\mu)}\left[R_{\ell+1}(c)\right]\right)\cdot\alpha^\vee=-1$,
for some simple root $\alpha\in S(X_r)$. This means that
$(w_{n-\ell}\nu^{n-\ell})\cdot\alpha^\vee=-1$ and that $r_\alpha
w_{n-\ell}\prec w_{n-\ell}$. 

By \dC\ then, the expression in \colbycol\ is contained in
\eqn\dwnl{\Theta\left( e^{\lambda+p_{R_\ell(\mu)}\left[R_\ell(c)\right]} 
d_{w_{n-\ell}}(e^{\mu-R_\ell(\mu)}) \right)\ ,} 
which vanishes, by \thdzero. The Bruhat order obeyed by the elements of
minimal defining chains, and their connection with Demazure characters,
ensures that all terms in \dwnl\ of the form \colbycol\ obeying \iplus\
are present in the left hand side of \Thi, if one is. 

A minimal defining chain $c$ is called $\lambda$-dominant if 
\eqn\lastd{\lambda+p_{R_i(\mu)}\left[R_i(c)\right]\in P_\ge (X_r)\ ,\ \ {\rm for\
all}\  0\le i\le n\ .}
So, only the $\lambda$-dominant MDCs can contribute to the left hand side of
\Thi. Define ${\cal C}_{\lambda,\mu}[\sigma]$ to be the set of
$\lambda$-dominant MDCs of $\mu$-weight $\sigma$. Then 
once the terms of the form \dwnl\ are
eliminated from the left the hand side of \Thi, we are left with 
\eqn\red{\Theta\left(\sum_{\nu\in P_\ge (X_r)} 
\dim {\cal C}_{\lambda,\mu}[\nu-\lambda]\ e^\nu\right)
\ =\
\Theta\left(\sum_{\nu\in P_\ge (X_r)} T_{\lambda,\mu}^\nu e^\nu\right)\ .}
But it is easy to see that
\eqn\Thsum{\eqalign{\Theta\left(\sum_{\mu\in P_\ge (X_r)} 
a_\mu e^\mu\right)
 =&\ 
\Theta\left(\sum_{\nu\in P_\ge (X_r)} b_\nu e^\nu\right)\ \ \cr 
\ \Rightarrow\ \ \ \ &a_\mu=b_\mu\ \ \ \ \ \forall\mu\in P_\ge (X_r)\ \ .\cr}}

Littelmann's generalization
of the Littlewood-Richardson rule then follows:
\eqn\Litt{\boxEq{T_{\lambda,\mu}^\nu\ =\ \vert{\cal
C}_{\lambda,\mu}[\nu-\lambda]\vert\ }\ .}

In the case $X_r=A_r$, the classical Littlewood-Richardson rule is recovered.
Let a standard tableau be called $\lambda$-dominant if the corresponding MDC
is $\lambda$-dominant. To find the tensor product coefficient
$T_{\lambda,\mu}^\nu$, one simply counts the number of
$\lambda$-dominant standard tableaux of shape $\mu$ and weight  $\nu-\lambda$.
These can be found by taking each standard tableau of shape $\mu$
and weight $\nu-\lambda$ and adding the weights of its columns, from right to
left, to  $\lambda$. If the addition of the weight of one of the columns
results in a non-dominant weight, no contribution results. If this does not
happen, the standard tableau contributes 1 to $T_{\lambda,\mu}^\nu$. 

The rule is most easily implemented by adding the columns directly to the
Young tableau of shape $\lambda$, to form a mixed tableau. The shape of the
mixed tableau is similar to the shape of a Young tableau. One takes the
columns of the standard tableau, and any box $\ST{\STrow{\b3}}$, e.g., to the
3rd row of the Young tableau, etc. If the addition in this manner of any column
results in a mixed tableau of non-dominant shape, the standard tableau does
not contribute. 

For example, consider the decomposition of the tensor product of $A_2$
representations: $R(2,0)\otimes R(2,1)=R(0,3)\oplus R(1,1)\oplus
R(2,2)\oplus R(3,0)\oplus R(4,1)$, where $R(\lambda_1,\lambda_2)$ means
$R(\lambda)$ with $\lambda=\lambda_1\omega^1 +\lambda_2\omega^2$. We have 
$T_{(2,0),(2,1)}^{(3,0)}=1$ while there are two standard tableaux of shape
$(2,1)$ and of the appropriate weight $(3,0)-(2,0)=(1,0)$:
\eqn\twoT{\ST{\STrow{\b1\b1\b2}\STrow{\b3}}\ ,\ \ 
\ST{\STrow{\b1\b1\b3}\STrow{\b2}}\ \ .}   
Only the first standard tableau contributes because  by adding its 
columns, from right to left, to the Young tableau $\ST{\STrow{\bv\bv}}$ 
of shape 
$(2,0)$, we obtain the following sequence of mixed tableaux:
\eqn\mTseq{\ST{\STrow{\bv\bv}\STrow{\b2}}\ ,\
\ST{\STrow{\bv\bv\b1}\STrow{\b2}}\ ,\ 
\ST{\STrow{\bv\bv\b1\b1}\STrow{\b2}\STrow{\b3}}\ ,}
each of which has dominant shape. In attempting to do the same for the second
standard tableau, we immediately encounter
\eqn\badmT{ \ST{ \STrow{\bv\bv}\vskip.38cm\STrow{\b3} }\ . }
This mixed
tableau is clearly not dominant, and so does not contribute to the tensor
product. 

Only five tableaux from Table 1 contribute to the tensor product
$R(2,0)\otimes R(2,1)$. The rest must result in terms of the form \thdzero.
For example, the five tableaux
\eqn\tpzero{ \ST{\STrow{\b1\b1\b3}\STrow{\b2}},\  
\ST{\STrow{\b1\b2\b3}\STrow{\b2}},\ 
\ST{\STrow{\b1\b1\b3}\STrow{\b3}},\ 
\ST{\STrow{\b1\b3\b3}\STrow{\b2}},\ 
\ST{\STrow{\b1\b3\b3}\STrow{\b3}},\ }
furnish together such a form, with $\kappa=(2,0)$, $\gamma=(2,1)$, and
$v=r_2r_1$. 

The form of the Littlewood-Richardson rule just described differs from the
original, in which the only numbered tableau that enters is 
\eqn\old{\ST{\STrow{\b1\b1\b1}\STrow{\b2}}\ \ .}
But the translation is simple, and is described explicitly in
\ref\Wey{Weyman, J.: Pieri's formulas for classical groups. Contemp. Math.
{\bf 88}, 177-184 (1989)}. The version involving standard tableaux is the
most convenient, since it can be directly related to vectors in
representations of $X_r$.


\newsec{\bf Modified Littlewood-Richardson Rules for Fusions}

In this section we will follow Littelmann's derivation of a generalized
Littlewood-Richardson rule \Lit, as outlined in the previous section, 
and attempt
to find a combinatorial rule for affine fusions. 

From \Vwzw\ and \Sch, we can follow steps in the previous section to arrive at 
\eqn\Thia{\Theta\left(e^\lambda{\rm ch}_\mu\right)(\sigma)\ =\
\Theta\left(\sum_{\nu\in P_\ge (X_{r,k})} 
N_{\lambda,\mu}^\nu e^\nu\right)(\sigma)\ ,
} where $\lambda,\mu,\sigma$ are elements of $P_\ge (X_{r,k})$. 

From \NT\ it
is clear that the Weyl group $\widehat W$ of $X_{r,k}$ must be implicated. 
Using the action \rzero\ of the $\widehat W$ generator $r_{\alpha_0}$ that is
additional to the generators of $W$, it is simple to show that 
\eqn\Thwa{\Theta\left(e^{w.\gamma}\right)(\sigma)\ =\ (\det w)\
\Theta\left(e^\gamma\right)(\sigma)\ ,\ \forall w\in {\widehat W}\ \ ,}
when $\sigma,\gamma\in P(X_r)$. But another ingredient in Littelmann's
derivation of a generalized Littlewood-Richardson rule is Demazure character
formulas for ${\rm ch}_\mu(\sigma)$ that involve the Bruhat order on $W$. We
know of no  way to write a similar formula for ${\rm ch}_\mu(\sigma)$ involving
the Bruhat order on the full affine Weyl group $\widehat W$. Formulas involving
the Bruhat order on subgroups of $\widehat W$ {\it can} be written, however, as
we now demonstrate. 

In order to use the affine Weyl symmetry \Thwa, the weights involved must have
level $k$: $\delta\cdot\gamma=k$ is required. If two such weights are added,
however, we get level $2k$. Let $\omega^0$ be the zeroth
affine fundamental weight, so that $\delta\cdot\omega^0=1$. If $\xi,\phi$
have level $k$, i.e. $\delta\cdot\phi=\delta\cdot\xi=k$, 
define $\bar\phi:=\phi-k\omega^0$. Then $\delta\cdot(\xi+\bar\phi)=k$. Notice
that $\sigma^{\omega^0}=1$, so that we are able to interpret the weight $\mu$ in
the left hand side of \Thia\ above as $\bar\mu$, in order to exploit the affine
symmetry \Thwa. 

With $w=r_{\alpha_0}$ in \Thwa, it is easy to show that
\eqn\Thdo{\Theta\left( d_{\alpha_0}(e^\gamma) \right)(\sigma)\ =\ 0,}
for $\delta\cdot\gamma=k$, with $\gamma\in P(X_r)$, if $d_{\alpha_0}$ is
defined by \dalpha\ with $\alpha=\alpha_0$. From this
\eqn\Thgdo{\Theta\left( e^\gamma d_{\alpha_0}(e^{\bar\phi})  
\right)(\sigma)\ =\ 0\ ,\ \ \ {\rm if}\ 
\gamma\cdot\alpha_0=k-\theta\cdot\gamma=0\ } follows, since
$\gamma+r_{\alpha_0}.\bar\phi=r_{\alpha_0}.(\gamma+\bar\phi)$ in that case.
Finally then, if we define Demazure operators $d_u$ for any  $u\in\widehat W$
by the reduced decomposition of $u$, we find \eqn\Thgdu{\Theta\left( e^\gamma
d_u(e^{\bar\phi})   \right)(\sigma)\ =\ 0\ ,\ \ \ {\rm if}\ 
k-\theta\cdot\gamma=0\ ,\ {\rm and}\ r_{\alpha_0}u\prec u\ ,}
where now $\prec$ indicates the Bruhat order on $\widehat W$.

For $X_r=A_r$, let $W^m$ denote the subgroup of $\widehat W$ generated by all
the affine primitive reflections except $r_{\alpha_m}$:
\eqn\Wm{W^m\ =\ \langle\ r_{\alpha_i}\ |\ i\in \{0,1,2,\ldots,r\}\backslash m\
\rangle\ .} So all $W^m\cong W$, and $W^0=W$. Denote by $x\in W$ the Weyl group
element that permutes the simple roots of $A_r$ and $-\theta$ in a cyclic manner:
\eqn\x{x(-\theta,\alpha_1,\alpha_2,\ldots,\alpha_r)\ =\ 
(\alpha_1,\alpha_2,\ldots,\alpha_r,-\theta)\ .}
Using these definitions we can write a modified Demazure character formula
\eqn\demaff{\boxEq{{\rm ch}_{\bar\mu}(\sigma)\ =\ \sum_{v\in W^m}\ d_v
(e^{x^m\bar\mu})(\sigma)}\ \ ,}
involving the subgroup $W^m$ of the affine Weyl group $\widehat W$.  
  
Defining chains and corresponding tableaux can also be associated to the
affine subgroups $W^m$. The highest weight $\mu$ of an integrable
representation $R(\mu)$ can be expressed uniquely as a sum
$\mu=x^m\nu^1+x^m\nu^2+\ldots +x^m\nu^n$, where $\nu^j\in F(A_r)$, and the
order $\nu^i\geq \nu^{i+1}$ has been fixed. A sequence 
$c=(w_1,w_2,\ldots,w_n)$ of elements of $W^m$ can be associated with such a
weight. If $w_j\preceq w_{j+1}$, the chain is a defining chain. If it is also
the minimal defining chain corresponding to the following sequence of weights:
\eqn\seqm{P^m_\mu(c)\ :=\
(w_1x^m\nu^1,w_2x^m\nu^2,\ldots,w_nx^m\nu^n)\ \ ,}
we will call it a $m$-MDC. 

Define the $\mu$-weight $p_{\mu,m}[c]$ of the $m$-MDC $c$ as
\eqn\pmum{p_{\mu,m}[c]\ =\
p_{\mu,m}[(w_1,w_2,\ldots,w_n)]\ :=\ w_1x^m\nu^1+\ldots+w_nx^m\nu^n\ .} 
Let ${\cal C}^m_\mu[\phi]$ be the set of $m$-MDCs
with $\mu$-weight $\phi$, and let ${\cal C}^m_\mu(w)$ denote the set of
$m$-MDCs with last element equal to $w\in W^m$. The weights of a representation
$R(\mu)$ and those of $m$-MDCs are related: \eqn\multcm{{\rm mult}_\mu(\phi)\ =\
\vert{\cal C}^m_\mu[\phi]\vert\ \ ,  }
so that 
\eqn\chcm{{\rm ch}_{\bar\mu}(\sigma)\ =\ \sum_{\phi\in P(\mu)}\ \sum_{c\in
{\cal C}^m_\mu[\phi]}\ e^{p_{\mu,m}[c]}(\sigma)\ \ .} 
The relevant connection with
Demazure operators is given by \eqn\dCm{d_w(e^\lambda)(\sigma)\ =\ \sum_{c\in
{\cal C}^m_\lambda(w)}\ e^{p_{\lambda,m}[c]}\ \ .}
The previous two equations are valid for any $\sigma\in P(A_r)$, and
the previous three are consequences of the Weyl invariance of characters.

A minor variation of the usual standard tableaux associated to $X_r=A_r$
can encode the $m$-MDCs. In terms of the orthonormal basis $\{e_i\}$ of
${\bf R}^{r+1}$ related to the usual standard tableaux, the action of $x\in
W$ is very simple:
\eqn\xei{x(e_1,e_2,\ldots,e_{r+1})\ =\ (e_2,e_3,\ldots,e_{r+1},e_1)\ .}
Using the same ordering of fundamental weights: $\omega^r>\omega^{r-1}>\cdots
>\omega^1$, we can define $m$-standard tableaux as the numbered tableaux of
shape $\mu$ with numbers appearing in the order
\eqn\order{(m+1,m+2,\ldots,r+1,1,2,\ldots,m)\ ,}
from left to right in its rows, and from top to bottom with no repetitions in
its columns.\foot{Other tableaux, with orderings that are non-cyclic
permutations of \order, can also be defined. Their relation to fusion
coefficients, however, is not clear.} Notice that 0-standard tableaux are the
usual standard tableaux. 

As an example, consider again the $A_2$ representation of highest weight
$2\omega^1+\omega^2=(2,1)$. For $A_2$, $x=r_{\alpha_1}r_{\alpha_2}$, and the
corresponding $1$-MDCs, and $1$-standard tableaux are shown in Table 2.  

\topinsert
\centerline{
{\vbox{\offinterlineskip
\hrule
\halign{& \vrule# & \strut \quad \hfill#\hfill \quad
& \vrule# & \strut \quad \hfill#\hfill \quad
& \vrule# & \strut \quad \hfill#\hfill \quad
& \vrule# & \strut \quad \hfill#\hfill \quad & \vrule# \cr
height3pt & \omit && \omit && \omit && \omit &\cr
& 1-MDC && weight\ sequence && tableau && weight  & \cr
height3pt & \omit && \omit && \omit && \omit &\cr
\noalign{\hrule}
height2pt & \omit && \omit && \omit && \omit &\cr
\noalign{\hrule}
height4pt & \omit && \omit && \omit && \omit &\cr
&$(id,id,id)$&&$((-1,0),(-1,1),(-1,1))$&&
$\ST{\STrow{\b2\b2\b2}\STrow{\b3}}$
&&$(-3,2)$ &\cr
height3pt & \omit && \omit && \omit && \omit &\cr
\noalign{\hrule}
height4pt & \omit && \omit && \omit && \omit &\cr
&$(r_0,r_0,r_0)$&&$((0,1),(-1,1),(-1,1))$&&
$\ST{\STrow{\b2\b2\b2}\STrow{\b1}}$
&&$(-2,3)$
&\cr height3pt & \omit && \omit && \omit && \omit &\cr
\noalign{\hrule}
height4pt & \omit && \omit && \omit && \omit &\cr
&$(id,id,r_2)$&&$((-1,0),(-1,1),(0,-1))$&&
$\ST{\STrow{\b2\b2\b3}\STrow{\b3}}$
&&$(-2,0)$
&\cr height3pt & \omit && \omit && \omit && \omit &\cr
\noalign{\hrule}
height4pt & \omit && \omit && \omit && \omit &\cr
&$(id,r_2,r_2)$&&$((-1,0),(0,-1),(0,-1))$&&
$\ST{\STrow{\b2\b3\b3}\STrow{\b3}}$
&&$(-1,-2)$
&\cr height3pt & \omit && \omit && \omit && \omit &\cr
\noalign{\hrule}
height4pt & \omit && \omit && \omit && \omit &\cr
&$(r_0,r_0,r_2r_0)$&&$((0,1),(-1,1),(0,-1))$&&
$\ST{\STrow{\b2\b2\b3}\STrow{\b1}}$
&&$(-1,1)$
&\cr height3pt & \omit && \omit && \omit && \omit &\cr
height4pt & \omit && \omit && \omit && \omit &\cr
&$(id,id,r_0r_2)$&&$((-1,0),(-1,1),(1,0))$&&
$\ST{\STrow{\b2\b2\b1}\STrow{\b3}}$
&&$(-1,1)$
&\cr height3pt & \omit && \omit && \omit && \omit &\cr
\noalign{\hrule}
height4pt & \omit && \omit && \omit && \omit &\cr
&$(id,r_2,r_0r_2)$&&$((-1,0),(0,-1),(1,0))$&&
$\ST{\STrow{\b2\b3\b1}\STrow{\b3}}$
&&$(0,-1)$
&\cr height3pt & \omit && \omit && \omit && \omit &\cr
height4pt & \omit && \omit && \omit && \omit &\cr
&$(r_0,r_2r_0,r_2r_0)$&&$((0,1),(0,-1),(0,-1))$&&
$\ST{\STrow{\b2\b3\b3}\STrow{\b1}}$
&&$(0,-1)$
&\cr height3pt & \omit && \omit && \omit && \omit &\cr
\noalign{\hrule}
height4pt & \omit && \omit && \omit && \omit &\cr
&$(r_2r_0,r_2r_0,r_2r_0)$&&$((1,-1),(0,-1),(0,-1))$&&
$\ST{\STrow{\b3\b3\b3}\STrow{\b1}}$
&&$(1,-3)$
&\cr height3pt & \omit && \omit && \omit && \omit &\cr
\noalign{\hrule}
height4pt & \omit && \omit && \omit && \omit &\cr
&$(r_0,r_0,r_0r_2)$&&$((0,1),(-1,1),(1,0))$&&
$\ST{\STrow{\b2\b2\b1}\STrow{\b1}}$
&&$(0,2)$
&\cr height3pt & \omit && \omit && \omit && \omit &\cr
\noalign{\hrule}
height4pt & \omit && \omit && \omit && \omit &\cr
&$(r_0,r_2r_0,r_0r_2r_0)$&&$((0,1),(0,-1),(1,0))$&&
$\ST{\STrow{\b2\b3\b1}\STrow{\b1}}$
&&$(1,0)$
&\cr height3pt & \omit && \omit && \omit && \omit &\cr
height4pt & \omit && \omit && \omit && \omit &\cr
&$(id,r_0r_2,r_0r_2)$&&$((-1,0),(1,0),(1,0))$&&
$\ST{\STrow{\b2\b1\b1}\STrow{\b3}}$
&&$(1,0)$
&\cr height3pt & \omit && \omit && \omit && \omit &\cr
\noalign{\hrule}
height4pt & \omit && \omit && \omit && \omit &\cr
&$(r_2r_0,r_2r_0,r_0r_2r_0)$&&$((1,-1),(0,-1),(1,0))$&&
$\ST{\STrow{\b3\b3\b1}\STrow{\b1}}$
&&$(2,-2)$
&\cr height3pt & \omit && \omit && \omit && \omit &\cr
\noalign{\hrule}
height4pt & \omit && \omit && \omit && \omit &\cr
&$(r_0,r_0r_2,r_0r_2)$&&$((0,1),(1,0),(1,0))$&&
$\ST{\STrow{\b2\b1\b1}\STrow{\b1}}$
&&$(2,1)$
&\cr height3pt & \omit && \omit && \omit && \omit &\cr
\noalign{\hrule}
height4pt & \omit && \omit && \omit && \omit &\cr
&$(r_2r_0,r_0r_2r_0,r_0r_2r_0)$&&$((1,-1),(0,-1),(0,-1))$&&
$\ST{\STrow{\b3\b1\b1}\STrow{\b1}}$
&&$(3,-1)$
&\cr height3pt & \omit && \omit && \omit && \omit &\cr
\noalign{\hrule}
}}}}

\smallskip
\leftskip=1cm
\rightskip=1cm
\noindent
\baselineskip=12pt
{\bf Table 2.} Minimal defining chains and corresponding 1-standard 
tableaux for the $A_2$ representation of highest weight
$2\omega^1+\omega^2=:(2,1),$ using the affine Weyl subgroup $W^1=<r_0,r_2>$.\bigskip
\leftskip=0cm
\rightskip=0cm
\baselineskip=13pt
\endinsert

The derivation of a modified Littlewood-Richardson rule from
\Thia,\Thgdu,\chcm\  and \dCm\ follows the derivation of the classical
Littlewood-Richardson rule outlined in the previous section. To state the
result, we define $c$, a $m$-MDC of length $n$, to be $(\lambda,k)$-dominant if 
\eqn\lakd{\lambda+p_{R_i(\mu)}\left[R_i(c)\right]\in P_\ge (X_{r,k})\ ,\ \ {\rm
for\ all}\  0\le i\le n\ .}
Let ${\cal C}^{m,k}_{\lambda,\mu}[\sigma]$ denote the set of
$(\lambda,k)$-dominant $m$-MDCs of $\mu$-weight $\sigma$. The modified
Littlewood-Richardson rule for $A_{r,k}$ fusions is then:
\eqn\Littm{\boxEq{N_{\lambda,\mu}^\nu\ \leq\ \vert{\cal
C}^{m,k}_{\lambda,\mu}[\nu-\lambda]\vert\ \ \forall\ 0\le m\le r\ .}}

Let a $m$-standard tableau be called $(\lambda,k)$-dominant if the corresponding
$m$-MDC is $(\lambda,k)$-dominant. To treat the fusion coefficient
$N_{\lambda,\mu}^\nu$ for $A_{r,k}$, one simply counts the number of
$(\lambda,k)$-dominant $m$-standard tableaux of shape $\mu$ and weight 
$\nu-\lambda$. These can be found by taking each $m$-standard tableau of shape
$\mu$ and weight $\nu-\lambda$ and adding the weights of its columns, from right
to left, to  $\lambda$. If the addition of the weight of one of the columns
results in a non-dominant weight $\not\in P_\geq(A_{r,k})$,
no contribution results. If this does not happen, the $m$-standard tableau
contributes 1 to the upper bound on $N_{\lambda,\mu}^\nu$. 

The modified rule is easily implemented by adding the columns directly to
the Young tableau of shape $\lambda$, to form a mixed tableau. If the addition
in this manner of any column results in a mixed tableau of shape
$\not\in P_\geq(A_{r,k})$, the $m$-standard tableau does not contribute. 
As an example, consider the $A_{2,k=3}$ fusion rule: $R(2,0)\otimes_3 R(2,1)
= R(1,1) \oplus R(0,3)$. Choosing $m=1$ as in Table 2, we can explain
$N_{(2,0),(2,1)}^{(3,0)}=0$ (for $k=3$). From Table 2 we read that the two
$1$-standard tableaux of shape $(2,1)$ and of the appropriate weight
$(3,0)-(2,0)=(1,0)$ are
\eqn\twomT{\ST{\STrow{\b2\b3\b1}\STrow{\b1}}\ ,\ \ 
\ST{\STrow{\b2\b1\b1}\STrow{\b3}}\ \ .}   
Neither of these tableaux contribute. The first does not, because 
by adding its 
columns, from right to left, to the Young tableau $\ST{\STrow{\bv\bv}}$ 
of shape
$(2,0)$, we obtain the following sequence of mixed tableaux:
\eqn\mmTseqi{\ST{\STrow{\bv\bv\b1}}\ ,\
\ST{\STrow{\bv\bv\b1}\vskip.38cm\STrow{\b3}}\ ,\ 
\ST{\STrow{\bv\bv\b1\b1}\STrow{\b2}\STrow{\b3}}\ ,}
the second of which has shape $\not\in P_\geq(A_{2,3})$. For the second
standard tableau, we find \eqn\mmTseqii{\ST{\STrow{\bv\bv\b1}}\ ,\
\ST{\STrow{\bv\bv\b1\b1}}\ ,\ 
\ST{\STrow{\bv\bv\b1\b1}\STrow{\b2}\STrow{\b3}}\ . }
Again, the second mixed tableau has a shape $\not\in P_\geq(A_{2,3})$, this
time because it has a number of columns (of less than 3 boxes) greater than
$k=3$. In a similar fashion, the complete fusion rule is found correctly, at
level 3 and all higher levels. 

If we compare the non-combinatorial expressions for tensor product coefficients
\Nmult, and for fusion coefficients \Tmult, it is clear that the modified
Littlewood-Richardson rule \Littm\ calculates the right hand side of 
\eqn\Tmmult{N^\nu_{\lambda,\mu}\ \le\ T^{(m) \nu}_{\la,\mu}\ :=\ \sum_{w\in
W^m}\ ({\rm det}w)\ {\rm mult}_\mu(w.\nu-\lambda)\ .}
We conjecture that 
\eqn\sat{\boxEq{N_{\lambda,\mu}^\nu\ =\ T^{(m) \nu}_{\lambda,\mu}\ \ {\rm when}\
\ {\rm mult}_\mu(r_{\alpha_m}.\nu-\lambda)=0}\ .}

It is not hard to find an example when the bound \Littm\ is not saturated.
Consider the $A_{4,4}$ fusion $R(1,1,1,1)\otimes_4 R(1,1,1,1)$, with
decomposition containing four copies of $R(1,1,1,1)$, so that
$N_{(1,1,1,1),(1,1,1,1)}^{(1,1,1,1)}=4$. The modified Littlewood-Richardson
rule gives $T_{(1,1,1,1),(1,1,1,1)}^{(m) (1,1,1,1)}=16$, 6, 5, 5, 6, for 
$m=$0, 1, 2, 3, 4, respectively. 

The $A_{4,4}$ fusion coefficient $N_{(1,1,1,1),(1,1,1,1)}^{(1,1,1,1)}=4$ 
{\it can} 
be recovered from the Littlewood-Richardson rule, however, if level-rank
duality\ref\KN{Kuniba, A., Nakanishi, T.: Level-rank
duality in fusion RSOS models. In:  Das, S.
et. al. (eds.): Modern Quantum Field Theory. New York: World Scientific
1991 (proccedings of the International
Colloquium on Modern  Quantum Field Theory, Bombay, 1990)}\ is used. But this
fix is not general. If the level is increased by one, it fails: the $A_{4,5}$ 
fusion coefficient $N_{(1,1,1,1),(1,1,1,1)}^{(1,1,1,1)}=14$ cannot be recovered
using \Littm\ and level-rank duality. The problem is deeper than that, as we
discuss in the next section. 

We now show that the modified Littlewood-Richardson rule \Littm\ gives a known 
upper bound on affine $A_r$ fusion
coefficients. Let
$J$ denote the diagram outer automorphism of $P_\ge(A_{r,k})$, with
action \eqn\tildeJ{ J \lambda\ =\  J\left(\sum_{i=1}^r
\lambda_i\omega^i \right)\ =\ \left( k-\sum_{i=1}^r\lambda_i \right)
\omega^1 + \sum_{i=2}^r \lambda_{i-1}\omega^i\ \ .}
What we
will show is that 
\eqn\ttm{T_{\lambda,\mu}^\nu\ =\ T_{\lambda,\mu}^{(0)\nu}\ =\  
T_{J^m\lambda,\mu}^{(m)J^m\nu}\ \ .}
From \Tmult\ we can write
\eqn\ttmi{T_{J^m\lambda,\mu}^{(m)J^m\nu}\ =\ \sum_{w\in W^m} (\det w)\ {\rm
mult}_\mu\left(J^m\left( (J^{-m}wJ^m)(\nu+\rho)-(\lambda+\rho)\right) \right)\
,}
using $J^m\rho=\rho$. But for $w\in W^m$, we have $J^{-m}wJ^m\in W$.
Furthermore, if $\sigma$ is any weight satisfying $\sigma\cdot\delta=0$, then
$J^m\sigma=x^m\sigma$. The Weyl invariance of the multiplicities then
establishes \ttm. 

The bound computed by \Littm\ is just \Tmmult, or
\eqn\NTJ{N_{\lambda,\mu}^\nu\ \leq\ T^{J^{-m}\nu}_{J^{-m}\lambda,\mu}\ .}
But it is well known that 
\eqn\NJN{N_{\lambda,\mu}^\nu\ =\ N^{J^{-m-n}\nu}_{J^{-m}\lambda,J^{-n}\mu}\ ,}
for all $m,n$. And with $N_{\lambda,\mu}^\nu\leq T_{\lambda,\mu}^\nu$, we
obtain 
\eqn\NTJJ{N_{\lambda,\mu}^\nu\ \leq\ T^{J^{-m-n}\nu}_{J^{-m}\lambda,J^{-n}\mu}\
.}
This last result is even stronger than the upper bound \Tmmult. It was
originally conjectured to be saturated for all triples $\lambda,\mu,\nu\in
P_\geq(A_{r,k})$, for some choice of $m$ and $n$ \FGP.

\newsec{\bf Discussion}

The classical Littlewood-Richardson rule for $A_r$ tensor products, and
generalizations for other simple Lie algebras (see \Litt), can be derived
from Demazure character formulas, as Littelmann demonstrated \Lit. Here we
wrote Demazure character formulas  \demaff\ for ratios of elements
of the modular $S$ matrices of affine Kac-Moody algebras.
These were then used in the Verlinde formula in an attempt to derive a
combinatorial rule for affine $A_r$ fusion rules. 

The resulting modified Littlewood-Richardson rule \Littm, however, only provides
an upper bound on the fusion coefficients. The basic reason is the use of the
subgroups $W^m$ of the affine Weyl group in the Demazure character formula
\demaff, instead of the full affine Weyl group and its Bruhat order. The
cancellations in \Tmult\ can be ordered systematically using the Bruhat order on
the finite Weyl group $W$, so that a combinatorial rule results from \dch\ and
\dC. The affine Bruhat order is needed, however, to order all the cancellations
in \Nmult\ in a similar way. But we were unable to write a character formula for
the relevant quantities \Sch\ that involves the Bruhat order on the full
affine Weyl group. 

Perhaps such a character formula can be written. On the other hand, a different 
derivation of the Littlewood-Richardson rule suggests it may be difficult. Let
$R(\mu;\phi)$ denote the subspace of the representation $R(\mu)$ consisting of
vectors of weight $\phi$. Then (see \ref\zelo{Zelobenko, D.P.: Compact Lie
groups and their representations. Providence, RI: American Mathematical
Society 1973}, Thm. 4, sect. 78, for example) 
\eqn\eqzelo{T_{\lambda,\mu}^\nu\ =\ \dim\left\{\ V\in R(\mu;\nu-\lambda)\
\vert\ E(-\alpha_i)^{\nu_i+1}V=0\ \forall i\in\{1,2,\ldots,r\}\ \right\}\ ,}
where $E(-\alpha_i)$ is the generator of $X_r$ in the Cartan-Weyl basis that
decreases the weight of a vector by $\alpha_i$. So-called ``good'' bases for
the representations of $X_r$ exist \ref\omath{Mathieu, O.: Good bases for
$G$-modules. Geometriae Dedicata {\bf 36}, 51-66 (1990)} that result in
combinatorial rules when used in \eqzelo. In particular, such a good basis for
$X_r=A_r$ can be labelled by the standard tableaux (or, equivalently, by
Gelfand-Tsetlin patterns \ref\gz{Gelfand, I.M., Zelevinsky, A.V.:
Multiplicities and proper bases for $g\ell(n)$. Group theoretical methods in
physics, proceedings of the third seminar, Yurmala. Moscow: Nauka 1985 (vol. 2,
pp. 22-31)}), so that the classical Littlewood-Richardson rule results. The
modified  Littlewood-Richardson rule
\Littm\  corresponds to 
\eqn\eqzelom{T_{\lambda,\mu}^{(m) \nu}\ =\ \dim\left\{\ V\in R(\mu;\nu-\lambda)\
\vert\ E(-\alpha_i)^{\nu_i+1}V=0\ \forall i\in\{0,1,2,\ldots,r\}\backslash m\
\right\}\ ,}
where we identify $E(-\alpha_0):=E(+\theta)$. A Weyl-transformed version of
the good basis of \gz\ exists, and the rule involving the $m$-standard
tableaux results. However, a natural generalization of the last two relations
is 
\eqn\eqzeloa{N_{\lambda,\mu}^\nu\ =\ \dim\left\{\ V\in R(\mu;\nu-\lambda)\
\vert\ E(-\alpha_i)^{\nu_i+1}V=0\ \forall i\in\{0,1,2,\ldots,r\}\ \right\}\ .}
This equation was conjectured in \ref\kmsw{Kirillov, A.N., Mathieu, P.,
S\'en\'echal, D., Walton, M.A.: Can fusion coefficients be calculated from the
depth rule? Nucl. Phys. {\bf B391}, 651-674 (1993)}, and was
motivated by the depth rule of \gw. But no basis of representations of $X_r$
exists that is good with respect to \eqzeloa\ \kmsw, as can be verified
explicitly for $X_r=A_r$ using the results of \gz.

It seems then that a different approach is required, one relating fusions
more directly to affine representations. This would allow the use of good
bases for affine representations, or formulas for their characters involving
the affine Bruhat order. In this context, we should mention the more recent
work of Littelmann: in \Lpaths\ the
generalized Littlewood-Richardson rule of \Littm\ was further generalized to a
rule for tensor products of all symmetrizable Kac-Moody algebras, and the
language of minimal defining chains was upgraded to one of certain paths.

\vskip 1truecm
\noindent{\it Acknowledgements}

Thanks go to C. Cummins for
conversations and for informing us of ref. \Lit, T. Gannon for helpful conversations 
and a 
critical reading of the manuscript, P. Mathieu for many helpful conversations,
and  V. Kac for correspondence
pointing out the problem of finding a combinatorial rule for affine fusions
(this was also mentioned in \GoodW). 

\listrefs
\bye